\documentclass[preprintnumbers,article,amsmath,amssymb,floatfix,10pt,prd,onecolumn,superscriptaddress,nofootinbib]{revtex4-2}

\usepackage{titlesec}
\usepackage[toc]{appendix}

\titleformat*{\section}{\LARGE\bfseries}
\titleformat*{\subsection}{\Large\bfseries}
\titleformat*{\subsubsection}{\large\bfseries}
\titleformat*{\paragraph}{\large\bfseries}
\titleformat*{\subparagraph}{\large\bfseries}
\usepackage{bm}
\usepackage{amsfonts}
\usepackage{latexsym}
\usepackage[latin1]{inputenc}
\usepackage{graphicx}
\usepackage{amsmath}
\usepackage{palatino}
\usepackage{mathpazo}
\usepackage{textcomp}
\usepackage{float}
\usepackage{booktabs}
\usepackage{dcolumn}
\usepackage{ragged2e}
\usepackage{hyperref}
\hypersetup{colorlinks,citecolor=blue}
\hypersetup{colorlinks=true,linkcolor=blue,filecolor=magenta,    urlcolor=blue}
\usepackage{amsthm}
\usepackage{xcolor}
\usepackage{orcidlink}
\usepackage{epsfig}
\usepackage[caption=false]{subfig}
\usepackage{commath}
\usepackage{cancel, ulem}

\usepackage{multirow}
\newcommand{\m}{\mathring}

\def\jnl@style{\it}
\def\aaref@jnl#1{{\jnl@style#1}}

\def\aaref@jnl#1{{\jnl@style#1}}

\def\aj{\aaref@jnl{AJ}}                   % Astronomical Journal
\def\apj{\aaref@jnl{ApJ}}                 % Astrophysical Journal
\def\apjl{\aaref@jnl{ApJ}}                % Astrophysical Journal, Letters
\def\apjs{\aaref@jnl{ApJS}}               % Astrophysical Journal, Supplement
\def\apss{\aaref@jnl{Ap\&SS}}             % Astrophysics and Space Science
\def\aap{\aaref@jnl{A\&A}}                % Astronomy and Astrophysics
\def\aapr{\aaref@jnl{A\&A~Rev.}}          % Astronomy and Astrophysics Reviews
\def\aaps{\aaref@jnl{A\&AS}}              % Astronomy and Astrophysics, Supplement
\def\mnras{\aaref@jnl{Mon.~Not.~Roy.~Astron.~Soc.}}             % Monthly Notices of the RAS
\def\prd{\aaref@jnl{Phys.~Rev.~D}}        % Physical Review D
\def\prc{\aaref@jnl{Phys.~Rev.~C}}  % Physical Review C
\def\prl{\aaref@jnl{Phys.~Rev.~Lett.}}    % Physical Review Letters
\def\qjras{\aaref@jnl{QJRAS}}             % Quarterly Journal of the RAS
\def\skytel{\aaref@jnl{S\&T}}             % Sky and Telescope
\def\ssr{\aaref@jnl{Space~Sci.~Rev.}}     % Space Science Reviews
\def\zap{\aaref@jnl{ZAp}}                 % Zeitschrift fuer Astrophysik
\def\nat{\aaref@jnl{Nature}}              % Nature
\def\aplett{\aaref@jnl{Astrophys.~Lett.}} % Astrophysics Letters
\def\apspr{\aaref@jnl{Astrophys.~Space~Phys.~Res.}} % Astrophysics Space Physics Research
\def\physrep{\aaref@jnl{Phys.~Rep.}}      % Physics Reports
\def\physscr{\aaref@jnl{Phys.~Scr}}       % Physica Scripta
\def\commat{\aaref@jnl{Comm.~Math.~Phys.}}              % Communications in Mathematical Physics
\def\science{\aaref@jnl{Science}}               % Science
\def\cqg{\aaref@jnl{Classical Quant.~Grav.}}            % Classical and Quantum Gravity
\def\jpcs{\aaref@jnl{JPCS}}                                     % Journal of Physics Conference Series
\def\ijmpd{\aaref@jnl{Int.~J.~Mod.~Phys.~D}}                    % International Journal of Modern Physics D
\def\grg{\aaref@jnl{Gen.~Relat.~Gravit.}}               % General Relativity and Gravitation
\def\rpp{\aaref@jnl{Rep.~Prog.~Phys.}}          % Reports on Progress in Physics
\def\npa{\aaref@jnl{Nucl.~Phys.~A}}        % Nuclear Physics A
\def\lrr{\aaref@jnl{Living Rev.~Rel.}}                   % Living reviews in relativity
\def\jcap{\aaref@jnl{J.~Cosmology Astropart.~Phys.}}    % Journal of cosmology and astroparticle physics
\def\rmp{\aaref@jnl{Rev.~Mod.~Phys.}}   %Reviews of modern physics
\def\epjc{\aaref@jnl{Eur.~Phys.~J.~C}} 
\def\plb{\aaref@jnl{~Phy.~Lett.~B}} 
\def\mpla{\aaref@jnl{Mod.~Phy.~Lett.~A}} 
\def\arxiv{\aaref@jnl{arxiv.org}}

\allowdisplaybreaks[1]
\begin{document}

\title{$f(Q,T)$ gravity, its covariant formulation, energy conservation and phase-space analysis}
\author{Tee-How Loo\orcidlink{0000-0003-4099-9843}}
\email{looth@um.edu.my}
\affiliation{Institute of Mathematical Sciences, Faculty of Science, Universiti Malaya, 50603 Kuala Lumpur, Malaysia}
\author{Raja Solanki\orcidlink{0000-0001-8849-7688}}
\email{rajasolanki8268@gmail.com}
\affiliation{Department of Mathematics, Birla Institute of Technology and
Science-Pilani,\\ Hyderabad Campus, Hyderabad-500078, India.}
\author{Avik De\orcidlink{0000-0001-6475-3085}}
\email{avikde@utar.edu.my}
\affiliation{Department of Mathematical and Actuarial Sciences, Universiti Tunku Abdul Rahman, Jalan Sungai Long,
43000 Cheras, Malaysia}
\author{P.K. Sahoo\orcidlink{0000-0003-2130-8832}}
\email{pksahoo@hyderabad.bits-pilani.ac.in}
\affiliation{Department of Mathematics, Birla Institute of Technology and
Science-Pilani,\\ Hyderabad Campus, Hyderabad-500078, India.}

%\date{}

%\footnotetext{The research was supported by the Ministry of Higher Education (MoHE), through the Fundamental Research Grant Scheme (FRGS/1/2021/STG06/UTAR/02/1). RS acknowledges UGC, New Delhi, India for providing Senior Research Fellowship with (UGC-Ref. No.: 191620096030) }

\begin{abstract}
 In the present article we analyze the matter-geometry coupled $f(Q,T)$ theory of gravity.
 %, newly proposed as an extension of the symmetric teleparallel theory, where $Q$ denotes the non-metricity scalar and $T$ the trace of energy-momentum tensor.
 We offer the fully covariant formulation of the theory, 
 %which produce an effective field equation comparable explicitly to general relativity (GR). Using this covariant formula of the field equation, 
 with which we %next 
 construct the correct energy balance equation and employ it to conduct a dynamical system analysis in a spatially flat Friedmann-Lema\^{i}tre-Robertson-Walker spacetime. We consider three different functional forms of the $f(Q,T)$ function, specifically, $f(Q,T)=\alpha Q+ \beta T$, $f(Q,T)=\alpha Q+ \beta T^2$, and $f(Q,T)=Q+ \alpha Q^2+ \beta T$ . We attempt to investigate the physical capabilities of these models to describe various cosmological epochs. We calculate Friedmann-like equations in each case and introduce some phase space variables to simplify the equations in more concise forms. We observe that the linear model $f(Q,T)=\alpha Q+ \beta T$ with $\beta=0$ is completely equivalent to the GR case without cosmological constant $\Lambda$. Further, we find that the model $f(Q,T)=\alpha Q+ \beta T^2$ with $\beta \neq 0$ successfully depicts the observed transition from decelerated phase to an accelerated phase of the universe. Lastly, we find that the model $f(Q,T)= Q+ \alpha Q^2+ \beta T$ with $\alpha \neq 0$ represents an accelerated de-Sitter epoch for the constraints $\beta < -1$ or $ \beta \geq 0$. 

\end{abstract}

%From the basis of our findings, we can conclude that both considered non-linear models can efficiently predict the de-Sitter type expansion of the universe and may represent a viable geometric alternative to dark energy.

\maketitle
%\footnotetext{   }

\tableofcontents
\section{Introduction}\label{sec1}
In the last two decades, the standard $\Lambda$CDM model, where the General Relativity has been manifested as the geometry of the background, provides an excellent fit to several cosmological observations. Nevertheless, the model is still struggling in the discrepancy of the value of the cosmological constant $\Lambda$. Thus, the standard theory of relativity may not be the ultimate theory to address the dark energy and the dark matter issues. This motivates a search for other theoretical cosmological scenarios that can describe cosmic acceleration with observational compatibility. General Relativity and its extensions, such as $f(R)$ gravity, $f(R,G)$ gravity, $f(R,T)$ gravity among others, were the curvature-based theory of gravitation formulated and extensively studied in the past \cite{cantata,modifiedgrav}. Another promising way to acquire the modified theory of gravitation is to formulate gravity on a flat spacetime geometry depending solely either on the torsion or the non-metricity, the first is known as metric teleparallel theories and the second as symmetric teleparallel theories. Nester and Yo \cite{Nester} proposed the latter and due to its dependence on the dark sector, Jimenez et al. extended it to formulate the $f(Q)$ gravity \cite{coincident} such that the late-time acceleration could be demonstrated from the additional geometric components. In the recent past, a tremendous amount of works were carried out in this $f(Q)$ theories \cite{Hohmann1,Hohmann2,FDA,jimenez1,jimenez2,lcdm,accfQ1,accfQ2,accfQ3,fQfT, fQfT1, fQfT3, gde, fQfT2,de/epjc,ad/ec, zhao, ad/bianchi,ad/ec, lin, cosmography, signa, red-shift,perturb, dynamical1}.

Very recently, a matter-geometry coupling in the form of $f(Q,T)$ theories were proposed \cite{Yixin/2019} in which the Lagrangian was represented by any viable function of the non-metricity scalar $Q$, and the trace $T$ of the energy-momentum tensor. Harko \cite{Harko} argued that this dependence can be caused by exotic imperfect fluids or quantum phenomena. Recently, several cosmological and astrophysical aspects of $f(Q,T)$ gravity have been tested, for instance, Energy conditions \cite{Arora1}, Baryogenesis \cite{BG}, Cosmological inflation \cite{Maryam}, Reconstruction of $f(Q,T)$ lagrangian \cite{Gadbail}, Static spherically symmetric wormholes \cite{Moreshwar}, Constraint on the effective equation of state \cite{Arora2}, Observational constraints on $f(Q,T)$ gravity models \cite{Arora3}, and Cosmological perturbations \cite{Antonio}. However, except the introductory article, most other studies were primarily conducted to contact observational datasets and not much theoretical investigation was carried out in this gravity theory which is still at its infancy at best. On a closer look, it is noticed that a covariant formulation was eluded so far, and there is some missing terms in the energy balance equation. 

This motivated us to derive the covariant formulation and the energy balance equation of the $f(Q,T)$ gravity. In addition, we present the asymptotic behavior of some cosmological $f(Q,T)$ models by utilizing the dynamical system technique.This approach is quite efficient in describing the asymptotic behavior of non-linear modified gravity models. Several cosmological models of the modified gravity have been tested by utilizing the dynamical system technique \cite{Mirza,Rudra,Leon1,Leon2,Khyllep}. One can investigate the asymptotic behavior of the cosmological model by analyzing the nature of critical points obtained by solving an autonomous system of first-order ordinary differential equations. The most important feature of any cosmological model is to have late-time stable solutions that depict the late-time behavior of the model. 

The manuscript is organized as follows: In Sec. \ref{sec2}, we present the mathematical formulation of $f(Q,T)$ gravity. Then in Sec. \ref{sec3}, we derive the covariant formula for $f(Q,T)$ gravity. In Sec. \ref{sec4}, we derive the corrected version of the energy-balance equation, both from the original metric field equation as well as from the novel covariant form of it. Further in Sec. \ref{sec5}, we investigate the asymptotic behavior of some cosmological $f(Q,T)$ models with the help of dynamical system analysis. Finally in Sec. \ref{sec6}, we summarize the obtained results.

%%%%%%%%%%%%%%%%%%%%%%%%%%%%%%%%%%%%%%%%%%%%%%%%%%%%%%%%%%%%%%%%%
%%%%%%%%%%%%%%%%%%%%%%%%%%%%%%%%%%%%%%%%%%%%%%%%%%%%%%%%%%%%%%%%%

\section{The mathematical formulation}\label{sec2}

In $f(Q,T)$-gravity theory, the background framework of the spacetime is the torsion free teleprallel geometry, i.e., $R^\rho{}_{\sigma\mu\nu} = 0$ and $T^\rho_{\mu\nu}=0$.
The difference between the associated connection $\Gamma^\lambda{}_{\mu\nu}$ and the Levi-Civita connection $\mathring{\Gamma}^\lambda{}_{\mu\nu}$ is known as the 
disformation tensor 
\begin{equation} \label{connc}
L^\lambda{}_{\mu\nu} = \Gamma^\lambda{}_{\mu\nu}- \m\Gamma^\lambda{}_{\mu\nu}\,,
\end{equation}
where $Q_{\lambda\mu\nu} := \nabla_\lambda g_{\mu\nu}$ is the non-metricity tensor. 
It follows that  
\begin{equation*}
L^\lambda{}_{\mu\nu} = \frac{1}{2} (Q^\lambda{}_{\mu\nu} - Q_\mu{}^\lambda{}_\nu - Q_\nu{}^\lambda{}_\mu) \,.
\end{equation*}
In addition, we define the non-metricity scalar 
\begin{equation} \label{Q}
Q = Q_{\lambda\mu\nu}P^{\lambda\mu\nu} = -\frac{1}{2}Q_{\lambda\mu\nu}L^{\lambda\mu\nu} + \frac{1}{4}Q_\lambda Q^\lambda - \frac{1}{2}Q_\lambda \tilde{Q}^\lambda \,,
\end{equation}
where
\begin{equation} \label{P}
P^\lambda{}_{\mu\nu} := \frac{1}{4} \left( -2 L^\lambda{}_{\mu\nu} + Q^\lambda g_{\mu\nu} - \tilde{Q}^\lambda g_{\mu\nu} -\frac{1}{2} \delta^\lambda_\mu Q_{\nu} - \frac{1}{2} \delta^\lambda_\nu Q_{\mu} \right) \,,
\end{equation}
is the superpotential tensor. Here we have used these two traces of non-metricity tensor:
 $Q_\lambda=Q_{\lambda\mu}{}^\mu$ and $\tilde Q_\lambda=Q_{\nu\lambda}{}^\nu$.
The action of $f(Q,T)$-gravity is given by
\begin{equation*}
S = \int \left[\frac{1}{2\kappa}f(Q,T) + \mathcal{L}_M \right] \sqrt{-g}\,d^4 x
\end{equation*}
where $g$ is the determinant of the metric tensor, $\mathcal{L}_M$ is the matter Lagrangian and
$T$ is the trace of the stress energy tensor $T_{\mu\nu}$, which is defined as 
\begin{align}
T_{\mu\nu}=-\frac 2{\sqrt{-g}}\frac{\delta(\sqrt{-g}\mathcal L_M)}{\delta g^{\mu\nu}}\,.
\end{align}
The variation of the action with respect to the metric, gives the metric field equation
\begin{equation} \label{FE1}
\frac{2}{\sqrt{-g}} \partial_\lambda (\sqrt{-g}f_QP^\lambda{}_{\mu\nu}) -\frac{1}{2}f g_{\mu\nu}
+f_T(T_{\mu\nu}+\Theta_{\mu\nu})
 + f_Q(P_{\nu\rho\sigma} Q_\mu{}^{\rho\sigma} -2P_{\rho\sigma\mu}Q^{\rho\sigma}{}_\nu) = \kappa T_{\mu\nu}\,,
\end{equation}
where $f_{(\cdot)}$ represents the partial derivative of $f$ with respect to $(\cdot)$ and 
\begin{align}
\Theta_{\mu\nu}=\frac{g_{\alpha\beta}\delta T_{\alpha\beta}}{\delta g^{\mu\nu}}\,.
\end{align}
Furthermore, varying the action with respect to the connection, we obtain the following connection field equation
\begin{align}\label{FE-connection}
\nabla_\mu\nabla_\nu(2\sqrt{-g}f_QP^{\mu\nu}{}_\lambda+\kappa H_\lambda{}^{\mu\nu})=0\,,
\end{align}
where 
\[
H_\lambda{}^{\mu\nu}=\frac{\sqrt{-g}}{2\kappa}f_T\frac{\delta T}{\delta\Gamma^\lambda{}_{\mu\nu}}
+\frac{\delta(\sqrt{-g}\mathcal L_M)}{\delta\Gamma^\lambda{}_{\mu\nu}}\,,
\]
is the hypermomentum tensor density.

Noticing that (\ref{FE1}) is valid only in the coincident gauge coordinate \cite{coincident}, it is essential to express it in its covariant form in the following section which firstly is independent of the choice  of coordinate systems and secondly provides an relatively straightforward manner in identifying the effective energy density and pressure. 

\section{Covariant Formulation of $f(Q,T)$}\label{sec3}
In the literature of cosmological application of symmetric teleparallel theory, we commonly observe the use of the so-called ``coincident gauge", that is, to identify a coordinate system whereby the connection vanishes %globally 
and covariant derivatives reduce to partial derivatives. This is common in both isotropic spatially flat FLRW spacetime as well as in Bianchi type anisotropic spacetimes. As described in \cite{zhao, covfT}, this sometime poses a major issue when we attempt to indulge into investigation of some other spacetimes using the same vanishing connection. Most of the time, the system is not consistent unless we force the non-metricity scalar $Q$ to be a constant, even worse, $Q=0$. To alleviate this problem, a fully-covariant formulation is very useful to employ non-vanishing connections into the game. In this section, we introduce the much-awaited covariant formulation of the $f(Q,T)$ theory.

Let us begin from the curvature free and torsion free constraints providing 
\begin{align}
\m R_{\mu\nu}+\m\nabla_\alpha L^\alpha{}_{\mu\nu}-\m\nabla_\nu\tilde L_\mu
+\tilde L_\alpha L^\alpha{}_{\mu\nu}-L_{\alpha\beta\nu}L^{\beta\alpha}{}_\mu=0\,,
\label{mRicci}\\
\m R+\m\nabla_\alpha(L^\alpha-\tilde L^\alpha)-Q=0\,. \label{mR}
\end{align}
As mentioned before, the coincident gauge is chosen in which  $\Gamma^\lambda{}_{\mu\nu}=0$ or $\m\Gamma^\lambda{}_{\mu\nu}=-L^\lambda{}_{\mu\nu}$.
Then we have   
\begin{align}\label{del:g}
\partial_\lambda\sqrt{-g}=-\sqrt{-g}\tilde L_\lambda\,.
\end{align}
Furthermore, we derive 
\begin{align}
\frac{2}{\sqrt{-g}}\partial_\lambda (\sqrt{-g}f_QP^\lambda{}_{\mu\nu})
+& f_Q(P_{\nu\rho\sigma} Q_\mu{}^{\rho\sigma} -2P_{\rho\sigma\mu}Q^{\rho\sigma}{}_\nu)
\notag\\
=&2(\nabla_\lambda f_Q)P^\lambda{}_{\mu\nu}
    +2f_Q(\m\nabla_\lambda P^\lambda{}_{\mu\nu}
    -L_{\alpha\beta\mu}P^{\beta\alpha}{}_\nu
    -L_{\alpha\beta\mu}P_\nu{}^{\beta\alpha}
    +L_{\nu\alpha\beta}P^{\alpha\beta}{}_\mu ) \notag\\
=&2(\nabla_\lambda f_Q)P^\lambda{}_{\mu\nu}
    +2f_Q(\m\nabla_\lambda P^\lambda{}_{\mu\nu}
    -\tilde L_\alpha L^\alpha{}_{\nu\mu}+L_{\alpha\beta\nu}L^{\beta\alpha}{}_\mu)\,.
    \label{I}
\end{align}
On the other hand, using (\ref{P}), (\ref{mRicci}) and (\ref{mR}),
we obtain
\begin{align}
2\m\nabla_\alpha P^\alpha{}_{\mu\nu}
=&-\m\nabla_\alpha L^\alpha{}_{\mu\nu}
    +\frac{\m\nabla_\alpha(L^\alpha-\tilde L^\alpha)}2g_{\mu\nu}
    +\frac{\m\nabla_\nu\tilde L_\mu+\m\nabla_\mu\tilde L^\nu}2 \notag\\
=&\m R_{\mu\nu} +\frac{Q-\m R}2g_{\mu\nu}
   +\tilde L_\alpha L^\alpha{}_{\mu\nu}-L_{\alpha\beta\nu}L^{\beta\alpha}{}_\mu\,.
   \label{II}
\end{align}
Combining (\ref{I})--(\ref{II}) we obtain
\begin{equation} \label{A}
\frac{2}{\sqrt{-g}} \partial_\lambda (\sqrt{-g}f_QP^\lambda{}_{\mu\nu}) 
 + f_Q(P_{\nu\rho\sigma} Q_\mu{}^{\rho\sigma} -2P_{\rho\sigma\mu}Q^{\rho\sigma}{}_\nu) =
 2(\nabla_\lambda f_Q)P^\lambda{}_{\mu\nu}+
f_Q \left(\m R_{\mu\nu} -\frac{\m R-Q}2g_{\mu\nu}\right)\,.
\end{equation}
Finally, the metric field equation can be rewritten covariantly as 
\begin{equation} \label{FE2}
f_Q \mathring{G}_{\mu\nu} + \frac{1}{2}g_{\mu\nu}(Qf_Q - f)
+f_T(T_{\mu\nu}+\Theta_{\mu\nu})
 %+ 2(f_{QQ} \nabla_\lambda Q+f_{QT}\nabla_\lambda T) P^\lambda{}_{\mu\nu} 
 + 2(\nabla_\lambda f_Q) P^\lambda{}_{\mu\nu} 
 = \kappa T_{\mu\nu}
\end{equation}
where $\m G_{\mu\nu}$ is the Einstein tensor corresponding to the Levi-Civita connection.
We define the effective stress energy tensor as
\begin{equation} \label{T^eff}
\kappa T^{\text{eff}}_{\mu\nu} = \kappa T_{\mu\nu}-f_T(T_{\mu\nu}+\Theta_{\mu\nu})-\frac{1}{2}g_{\mu\nu}(Qf_Q-f)
 -2(f_{QQ} \nabla_\lambda Q+f_{QT}\nabla_\lambda T) P^\lambda{}_{\mu\nu} \,.
\end{equation}
In the present paper, we consider a perfect fluid type spacetime, whose stress energy tensor takes the form 
\begin{align}
T_{\mu\nu}=pg_{\mu\nu}+(p+\rho)u_\mu u_\nu
\end{align}
to which the matter Lagrangian can be taken as $\mathcal L_M=p$, where $\rho$, $p$ and $u^\mu$ denote 
the energy density, pressure and four velocity of the fluid respectively. It follows that 
\begin{align}
\Theta_{\mu\nu}=pg_{\mu\nu}-2T_{\mu\nu}
\end{align}

%%%%%%%%%%%%%%%%%%%%%%%%%%%%%%%%%%%%%%%%%%%%%%%%%%%%%%%%%%%%%%%%%
%%%%%%%%%%%%%%%%%%%%%%%%%%%%%%%%%%%%%%%%%%%%%%%%%%%%%%%%%%%%%%%%%
\section{Energy conservation in $f(Q,T)$ theory}\label{sec4}
The $f(Q,T)$ theory is not compatible with the energy conservation criterion, so an energy momentum balance equation was offered in \cite{Yixin/2019}. Unfortunately, it seems to be not derived correctly as two crucial terms were noticed missing, particularly while applying the covariant derivative $\m\nabla_\mu$ to
\begin{align*}
\frac2{\sqrt{-g}}\nabla_\alpha(\sqrt{-g}f_QP^{\alpha\mu}{}_\nu) \qquad \text{  and  } \qquad f_TT^\mu{}_\nu.
\end{align*}
The corrected energy-momentum balance equation should read as 
\begin{align}\label{eqn:balance}
\m\nabla_\mu T^\mu{}_\nu=\frac1{\kappa-f_T}\left\{
\m\nabla_\mu(f_T\Theta^\mu{}_\nu)+(\nabla_\mu f_T)T^\mu{}_\nu-\frac12f_T\nabla_\nu T-\frac\kappa{\sqrt{-g}}\nabla_\alpha\nabla_\mu H_\nu{}^{\alpha\mu}
\right\}\,.
\end{align}
In what follows, we briefly discuss the derivation of (\ref{eqn:balance}). Let us begin with
the field equation (\ref{FE1}) of type $(1,1)$ %can be rewritten as 
\begin{align}
 \label{FE1b}
 \kappa T^\mu{}_\nu-f_T(\Theta^\mu{}_\nu+T^\mu{}_\nu)+\frac12f\delta^\mu{}_\nu
 =f_QP^{\mu\rho\sigma} Q_{\nu\rho\sigma}
 +\frac{2}{\sqrt{-g}}\nabla_\lambda (\sqrt{-g}f_QP^{\lambda\mu}{}_\nu) \,. 
\end{align}
The divergence of the preceding equation gives 
\begin{align}\label{eqn:divT-b}
(\kappa-f_T)\m\nabla_\mu T^\mu{}_\nu-(\nabla_\mu f_T)T^\mu{}_\nu
-\m\nabla_\nu(f_T\Theta^\mu{}_\nu)+\frac12\nabla_\nu f
&
=(\nabla_\lambda f_Q)P^{\mu\rho\sigma} Q_{\nu\rho\sigma})
 +\m\nabla_\mu\left\{\frac{2}{\sqrt{-g}}\nabla_\lambda (\sqrt{-g}f_QP^{\lambda\mu}{}_\nu) \right\}\,.
\end{align}
To simplify the above equation, we explicitly expand the two terms in the right hand side as 
\begin{align}
 \m\nabla_\mu(f_QP^{\mu\rho\sigma} Q_{\nu\rho\sigma})
 =&Q_{\nu\rho\sigma}(\nabla_\mu-\tilde L_\mu)(f_QP^{\mu\rho\sigma})
 +f_QP^{\mu\rho\sigma}(\nabla_\mu Q_{\nu\rho\sigma}
 +L^\beta{}_{\nu\mu}Q_{\beta\rho\sigma}) \nonumber\\
 =&-2L_{\sigma\nu\rho}(\nabla_\mu-\tilde L_\mu)(f_QP^{\mu\rho\sigma})
 +f_QP^{\mu\rho\sigma}(\m\nabla_\nu Q_{\mu\rho\sigma}
 -2L^\beta{}_{\rho\nu}Q_{\mu\beta\sigma})\,,
\end{align}
\begin{align}
\m\nabla_\mu\left\{
\frac{2}{\sqrt{-g}}\nabla_\lambda (\sqrt{-g}f_QP^{\lambda\mu}{}_\nu) 
\right\}
=&(\nabla_\mu-\tilde L_\mu)\left\{
\frac{2}{\sqrt{-g}}\nabla_\lambda (\sqrt{-g}f_QP^{\lambda\mu}{}_\nu) 
\right\}+
\frac{2L^\beta{}_{\nu\mu}}{\sqrt{-g}}\nabla_\lambda (\sqrt{-g}f_QP^{\lambda\mu}{}_\beta) \nonumber\\
=&\frac{2}{\sqrt{-g}}\nabla_\mu\nabla_\lambda (\sqrt{-g}f_QP^{\lambda\mu}{}_\nu) 
+
2L^\beta{}_{\nu\rho}
(\nabla_\mu-\tilde L\mu)(f_QP^{\mu\rho}{}_\beta) 
\,,
\end{align}
where we have used the relation (\ref{del:g}).
%\begin{align}
%\partial_\mu\frac1{\sqrt{-g}}=\frac{\tilde L_\mu}{\sqrt{-g}}\,.
%\end{align}
In addition, we have the relations
\begin{align}\label{ad1}
2L^\beta{}_{\nu\rho}(\nabla_\mu-\tilde L\mu)(f_QP^{\mu\rho}{}_\beta) 
=L_{\sigma\nu\rho}(\nabla_\mu-\tilde L\mu)(f_QP^{\mu\rho\sigma}) 
 +L^\beta{}_{\nu\rho}(f_QP^{\mu\rho}{}_\beta)Q_{\mu\sigma\beta}
\end{align}
\begin{align}\label{ad2}
\nabla_\nu Q=\m\nabla_\nu(P^{\mu\rho\sigma}Q_{\mu\rho\sigma})
=2P^{\mu\rho\sigma}\m\nabla_\nu Q_{\mu\rho\sigma}\,.
\end{align}
Finally, (\ref{eqn:balance}) can be obtained after 
substituting (\ref{FE-connection}) and the above relations into (\ref{eqn:divT-b}).

Now that we offer the fully covarinat formulation of the field equation (\ref{FE2}), we can directly use the Bianchi identity to derive another equivalent form of the energy momentum balance equation by taking divergence of (\ref{FE2})
\begin{align}\label{eqn:divT-c}
(\kappa-f_T)\m\nabla_\mu T^\mu{}_\nu
-&(\nabla_\mu f_T)T^\mu{}_\nu
-\m\nabla_\nu(f_T\Theta^\mu{}_\nu)+\frac{f_T}2\nabla_\nu Q \nonumber\\
=&(\nabla_\lambda f_Q)\left(\m G^\lambda{}_\nu+\frac Q2\delta^\lambda{}_\nu+2\m\nabla_\mu P^{\lambda\mu}{}_\nu\right)
 +2(\m\nabla_\mu\m\nabla_\lambda f_Q)P^{\lambda\mu}{}_\nu\,.
\end{align}
Note that (\ref{eqn:balance}) and (\ref{eqn:divT-c}) of the energy-balance in $f(Q,T)$ theory are identical once the affine connection field equation (\ref{FE-connection}) are taken into account.
%\begin{align}
%\m\nabla_\mu T^\mu{}_\nu=\frac1{\kappa-f_T}\left\{\m\nabla_\mu(f_T\Theta^\mu{}_\nu)+(\nabla_\mu f_T)T^\mu{}_\nu-\frac12f_T\nabla_\nu T+\frac2{\sqrt{-g}}\nabla_\mu\nabla_\nu(\sqrt{-g}f_QP^{\mu\nu}{}_\lambda)\right\}\,.\end{align}

%%%%%%%%%%%%%%%%%%%%%%%%%%%%%%%%%%%%%%%%%%%%%%%%%%%%%%%%%%%%%%%%%
%%%%%%%%%%%%%%%%%%%%%%%%%%%%%%%%%%%%%%%%%%%%%%%%%%%%%%%%%%%%%%%%%
%\section{Equations of motion}\label{sec4}

%%%%%%%%%%%%%%%%%%%%%%%%%%%%%%%%%%%%%%%%%%%%%%%%%%%%%%%%%%%%%%%%%
%%%%%%%%%%%%%%%%%%%%%%%%%%%%%%%%%%%%%%%%%%%%%%%%%%%%%%%%%%%%%%%%%

\section{Cosmological application}\label{sec5}

In this section we explore some cosmological applications of the previously obtained covariant formulation and energy balance equation. For this purpose we consider the spatially flat homogeneous and isotropic FLRW spacetime with the line element given in Cartesian coordinates by
\begin{align}\label{metric}
    ds^2=-dt^2+a^2(t)\left[dx^2+dy^2+dz^2\right].
\end{align}
For simplicity, we utilise the usual vanishing affine connection in this coincident gauge choice to obtain the non-metricity scalar as
\begin{align}
    Q=-6H^2.
\end{align}
Using \eqref{FE2} and \eqref{metric}, we obtain the following Friedmann-like equations.
\begin{align}
(\kappa +f_T)\rho+f_T p &= \frac{f}{2} + 6H^2 f_Q,  \label{rho-temp}\\
%\end{align}
%From (\ref{rho})--(\ref{p}), we obtain
%\begin{align}
\kappa p &= -\frac{f}{2} -6H^2 f_Q - \frac{\partial}{\partial t}(2Hf_Q)\,.
    \label{p}
\end{align}
When particularly $f_T=-\kappa$, we obtain  the equation
\begin{align}
\frac{\partial}{\partial t}(2Hf_Q)=0\,,
\end{align}
whose solution is given as  
\begin{align}
f(Q,T)=\alpha\sqrt{-Q}-\kappa T\,,
\end{align}
where $\alpha$ is a constant. 
As this solution does not include GR, we conclude that this is not an adequate model. In what follows, we consider only $f_T \neq-\kappa$. Using (\ref{rho-temp})--(\ref{p}), we obtain
\begin{align}
\kappa\rho &=\frac f2+6H^2f_Q+\frac{f_T}{\kappa+f_T} \frac{\partial}{\partial t}(2Hf_Q)\,.
        \label{rho}
%(\kappa+f_T)(\rho+p)=&- \frac{\partial}{\partial t}(2Hf_Q). \label{rho+p}
\end{align}
From the corrected energy balance equation (\ref{eqn:divT-c}) derived in the last section, we can write the continuity relation as  
\begin{align}\label{gen}
-\dot\rho-3H(\rho+p)
=-\frac1\kappa\frac{\partial}{\partial t}\left(\frac f2+6H^2f_Q\right)
-\frac1{\kappa+f_T}\left(\frac{\dot f_T}{\kappa+f_T}-3H\right)\frac{\partial}{\partial t}(2Hf_Q)
  -\frac1\kappa\frac{f_T}{\kappa+f_T}\frac{\partial^2}{\partial t^2}(2Hf_Q).
\end{align}
In particular, for pressureless dust era ($p=0$) it reduces to
\begin{align}
\dot\rho+3H\rho
=\frac1{\kappa+f_T}
 \left\{\frac{\dot f}2+12\dot HHf_Q+6H^2\dot{f_Q}
+\left(3H-\frac{\dot f_T}{\kappa+f_T}\right)\left(\frac f2+6H^2f_Q\right)
\right\}\,.\label{dust}
\end{align}

On the other hand, we can derive the effective energy density and pressure equations using (\ref{T^eff})
 \begin{align}
\kappa \rho^{\text{eff}}=&(\kappa +f_T)\rho+f_T p- \frac{f}{2} - 3H^2 f_Q\,,  \\
 \kappa p^{\text{eff}} 
=&\kappa p +\frac{f}{2} + 3H^2f_Q+2H\dot{f}_Q 
\,.
\end{align}

Now that we pull the necessary things together, in the following subsections we are going to examine the dynamical behavior of some well known cosmological $f(Q,T)$ models by incorporating the phase space approach.

\subsection{Linear $f(Q,T)$ model}
We consider the following linear $f(Q,T)$ model
\begin{equation}\label{a0}
f(Q,T)= \alpha Q+ \beta T 
\end{equation}
Here $\alpha$ and $\beta$ are free model parameters. The cosmological implications of the considered model have been investigated in \cite{Yixin/2019}. We are going to investigate the asymptotic behavior of the model. 
The Friedmann equations \eqref{p} and \eqref{rho} corresponding to our linear $f(Q,T)$ model, for the dust case, becomes
\begin{equation}\label{a1}
    (2\kappa+3\beta) \rho = 6 \alpha H^2
\end{equation}
\begin{equation}\label{a2}
    2 \dot{H} + 3 H^2 = \frac{\beta}{2\alpha} \rho\,,
\end{equation}
and the equation \eqref{dust} becomes
\begin{equation}\label{a3}
 \dot{\rho}  + \frac{6(\kappa+\beta)}{(2\kappa+3\beta)} H \rho = 0
\end{equation}
We define the following phase space variables 
\begin{equation}
    x=\frac{ (2\kappa+3\beta) \rho}{6 \alpha H^2}  \:\: \text{and} \:\: y= \frac{1}{\frac{H_0}{H}+1}
\end{equation}\label{a4}
Then we have constraints $x=1$ and $0 \leq y \leq 1 $ .
Now we define $N=dln(a)$, then corresponding to our linear $f(Q,T)$ model we obtained the following autonomous system by using \eqref{a1}-\eqref{a3},
\begin{equation}\label{a5}
    x'= \frac{dx}{dN} = \frac{3\beta}{(2\kappa+3\beta)} x (1-x)
\end{equation}
\begin{equation}\label{a6}
    y'= \frac{dy}{dN} = \frac{3}{2} \big[ \frac{\beta x }{(2\kappa+3\beta)} -1 \big] y (1-y)
\end{equation} 

Now by the definition of deceleration and EoS parameter, we have for $x=1$
\begin{equation}\label{a8}
    q=-1+ \frac{3(\kappa+\beta)}{(2\kappa+3\beta)} \: \text{and} \: \omega = -1+ \frac{2(\kappa+\beta)}{(2\kappa+3\beta)} 
\end{equation}

The nature of the critical points obtained by solving the above autonomous equations with $\kappa=1$, are presented below in the Table \eqref{Table-1}.

\begin{table}[H]
\begin{center}\caption{Table shows the critical points and their behavior corresponding to the model $f(Q,T)=\alpha Q+ \beta T$.}
\begin{tabular}{|c|c|c|c|c|c|}
\hline
 Critical Points $(x_c,y_c)$ & Eigenvalues $\lambda_1$ and $\lambda_2$ & Nature of critical point  & $q$ & $\omega$ \\
\hline 
 $P(1,1)$ & $ -\frac{3 \beta }{(2+3 \beta) } \:\: \text{and} \:\: \frac{3 (1+\beta )}{(2+3 \beta) }$ & Stable for $-1<\beta <-\frac{2}{3}$ & $\frac{1}{2+3\beta}$ & $-\frac{\beta}{2+3\beta}$ \\
$Q(1,0)$ & $-\frac{3 (1+\beta )}{(2+3 \beta )} \:\: \text{and} \:\: -\frac{3 \beta }{(2+3 \beta) }$ & Stable for $\beta < -1$ or $ \beta \geq 0$ & $\frac{1}{2+3\beta}$ & $-\frac{\beta}{2+3\beta}$ \\
\hline
\end{tabular}\label{Table-1}
\end{center}
\end{table}

The asymptotic behaviour of our linear $f(Q,T)$ model corresponding to the case $\beta=0$  with $\kappa=1$, presented below in Fig. \eqref{f1}.

\begin{figure}[H]
\begin{center}
\includegraphics[scale=0.585]{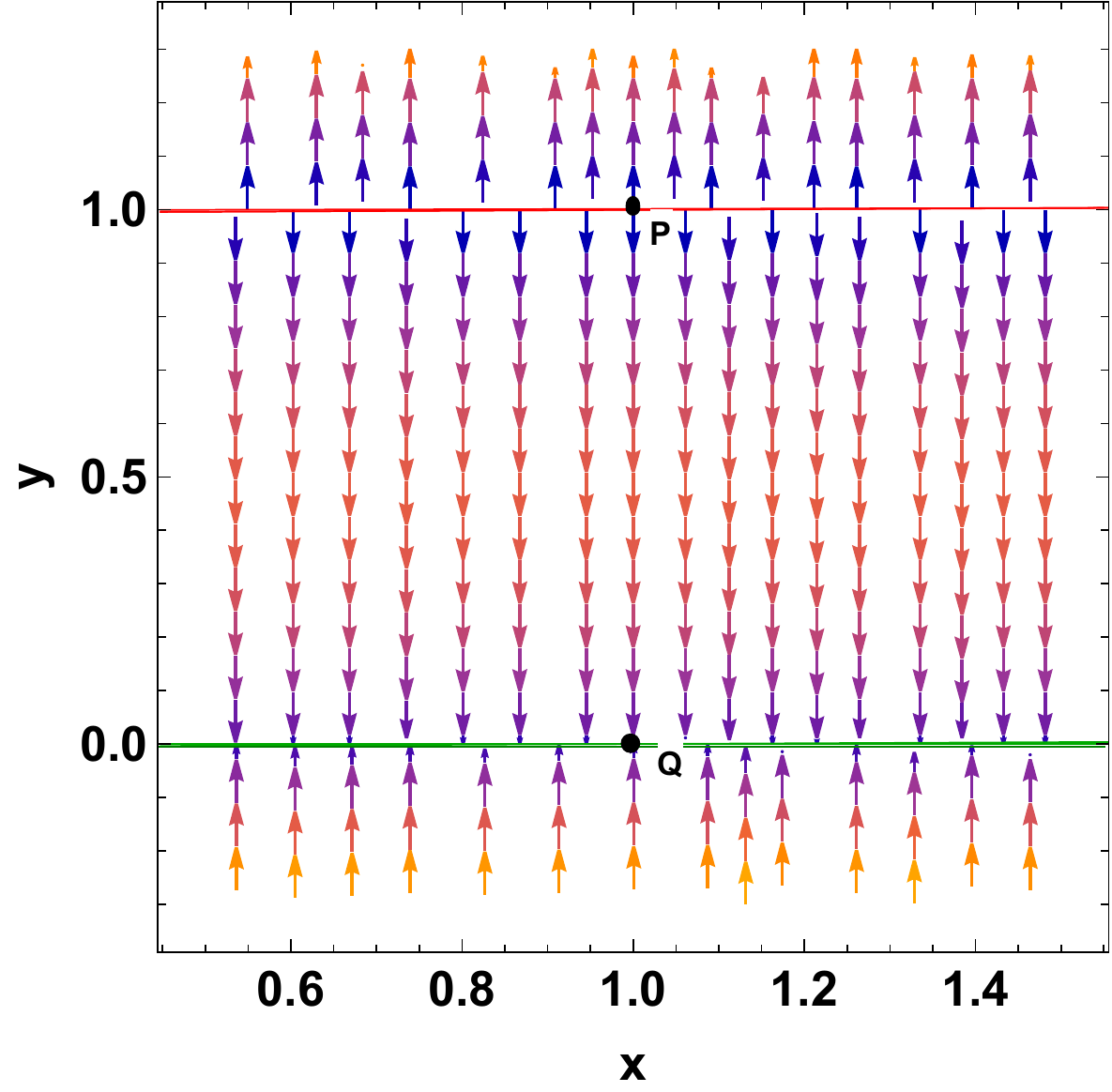}
\caption{Phase plots corresponding to the case $\beta=0$ with $\kappa=1$.}\label{f1}
\end{center}
\end{figure}

For the case $\beta=0$, the obtained critical points are P(1,1) and Q(1,0) with corresponding eigenvalues $0, \frac{3}{2}$ and $0, -\frac{3}{2}$ respectively. From Fig. \eqref{f1}, it is evident that the trajectories are emerging from the unstable past attractor P(1,1) and converging to future attractor Q(1,0). Hence, point P(1,1) is a source, whereas point Q(1,0) is a sink. From equation \eqref{a8}, we obtained $q=\frac{1}{2}$ and $\omega=0$ for both P and Q. Therefore, we can conclude that our model $f(Q,T)=\alpha Q+ \beta T$ with $\beta=0$ remains lies in the matter dominated epoch and cannot describe the accelerated era as well as the initial singularity. Hence, it is completely equivalent to the GR case without cosmological constant $\Lambda$.  \\

%\textbf{Case II :}
%For the case $\beta=-1$, we obtained non-isolated critical points R(1,y) with eigenvalues $0$ and $-3$. From the right panel of Fig. \eqref{f1}, it is evident that the trajectories are converging to the line of critical points R(1,y). Hence, R(1,y), for any y, is a sink. From equation \eqref{a8}, we obtained $q=-1$ and $\omega=-1$. Therefore we can conclude that our model $f(Q,T)=\alpha Q+ \beta T$ with $\beta=-1$ remains in the accelerated de-Sitter epoch throughout the evolution.  \\

\justify From the equation \eqref{a8}, one can notice that the linear model $f(Q,T)=\alpha Q + \beta T$ cannot describes the transition from decelerated to accelerated era for any parameter values.

\subsection{Non-Linear $f(Q,T)$ model}
\subsubsection{Model-I}

We consider the following non-linear $f(Q,T)$ model
\begin{equation}\label{b0}
f(Q,T)= \alpha Q+ \beta T^2 
\end{equation}
Here $\alpha$ and $\beta \neq 0$ are free model parameters. The cosmological implications of the considered model have been investigated in \cite{Yixin/2019}. We are going to investigate the asymptotic behavior of the model. 
The Friedmann equations \eqref{p} and \eqref{rho} corresponding to our non-linear $f(Q,T)$ model, for the dust case, becomes
\begin{equation}\label{b1}
    (2\kappa-5\beta \rho ) \rho = 6 \alpha H^2 
\end{equation}
\begin{equation}\label{b2}
    2 \dot{H} + 3 H^2  = -\frac{\beta}{2 \alpha} \rho^2
\end{equation}
and the equation \eqref{dust} becomes
\begin{equation}\label{b3}
 \dot{\rho}  + \frac{3(\kappa-2\beta \rho)}{(\kappa-5\beta \rho)}  H \rho = 0
\end{equation}

We define the following phase space variables 
\begin{equation}\label{b4}
    x=\frac{\kappa \rho}{3 \alpha H^2},  \:\: y= \frac{1}{\frac{H_0}{H}+1} \:\:  \text{and} \:\: z= -\frac{5 \beta \rho^2}{6 \alpha H^2}
\end{equation}\label{a4}
Then we have constraints $x+z=1$ and $0 \leq y \leq 1 $ .
Now corresponding to our non-linear $f(Q,T)$ model we obtained the following autonomous system with respect to the variable $N=dln(a)$, by using \eqref{b1}-\eqref{b3} and the constrain $x+z=1$,
\begin{equation}\label{b5}
    x'= \frac{dx}{dN} = \frac{3x (x-1)(x+4)}{5(x-2)} 
\end{equation}
\begin{equation}\label{b6}
    y'= \frac{dy}{dN} = \frac{3}{10} y (y-1) (x+4)
\end{equation} 

Now by the definition of deceleration and EoS parameter, we have 
\begin{equation}\label{b8}
    q= \frac{1}{10} (3x+2) \: \text{and} \: \omega = \frac{1}{5} (x-1)
\end{equation}

The nature of the critical points obtained by solving the above autonomous equations with $\kappa=1$, are presented below in the Table \eqref{Table-2}.

\begin{table}[H]
\begin{center}\caption{Table shows the critical points and their behavior corresponding to the model $f(Q,T)=\alpha Q+ \beta T^2$.}
\begin{tabular}{|c|c|c|c|c|}
\hline
 Critical Points $(x_c,y_c)$ & Eigenvalues $\lambda_1$ and $\lambda_2$ & Nature of critical point  & $q$ & $\omega$ \\
\hline 
 $A(0,0)$ & $-\frac{6}{5} \:\: \text{and} \:\: \frac{6}{5}$ & Saddle & $\frac{1}{5}$ & $-\frac{1}{5}$ \\
 \hline 
 $B(1,1)$ & $-3 \:\: \text{and} \:\: \frac{3}{2}$ & Saddle & $\frac{1}{2}$ & $0$ \\
\hline
 $C(1,0)$  & $-3 \:\: \text{and} \:\: -\frac{3}{2}$ & Stable & $\frac{1}{2}$ & $0$ \\
\hline
 $D(0,1)$  & $\frac{6}{5} \:\: \text{and} \:\: \frac{6}{5}$ & Unstable & $\frac{1}{5}$ & $-\frac{1}{5}$ \\
 \hline
 $E(-4,y)$  & $-2 \:\: \text{and} \:\: 0$ & Stable & $-1$ & $-1$ \\
 \hline 
\end{tabular}\label{Table-2}
\end{center}
\end{table}

The asymptotic behaviour of our non-linear $f(Q,T)$ model with $\kappa=1$, presented below in Fig. \eqref{f2}.

\begin{figure}[H]
\begin{center}
\includegraphics[scale=0.6]{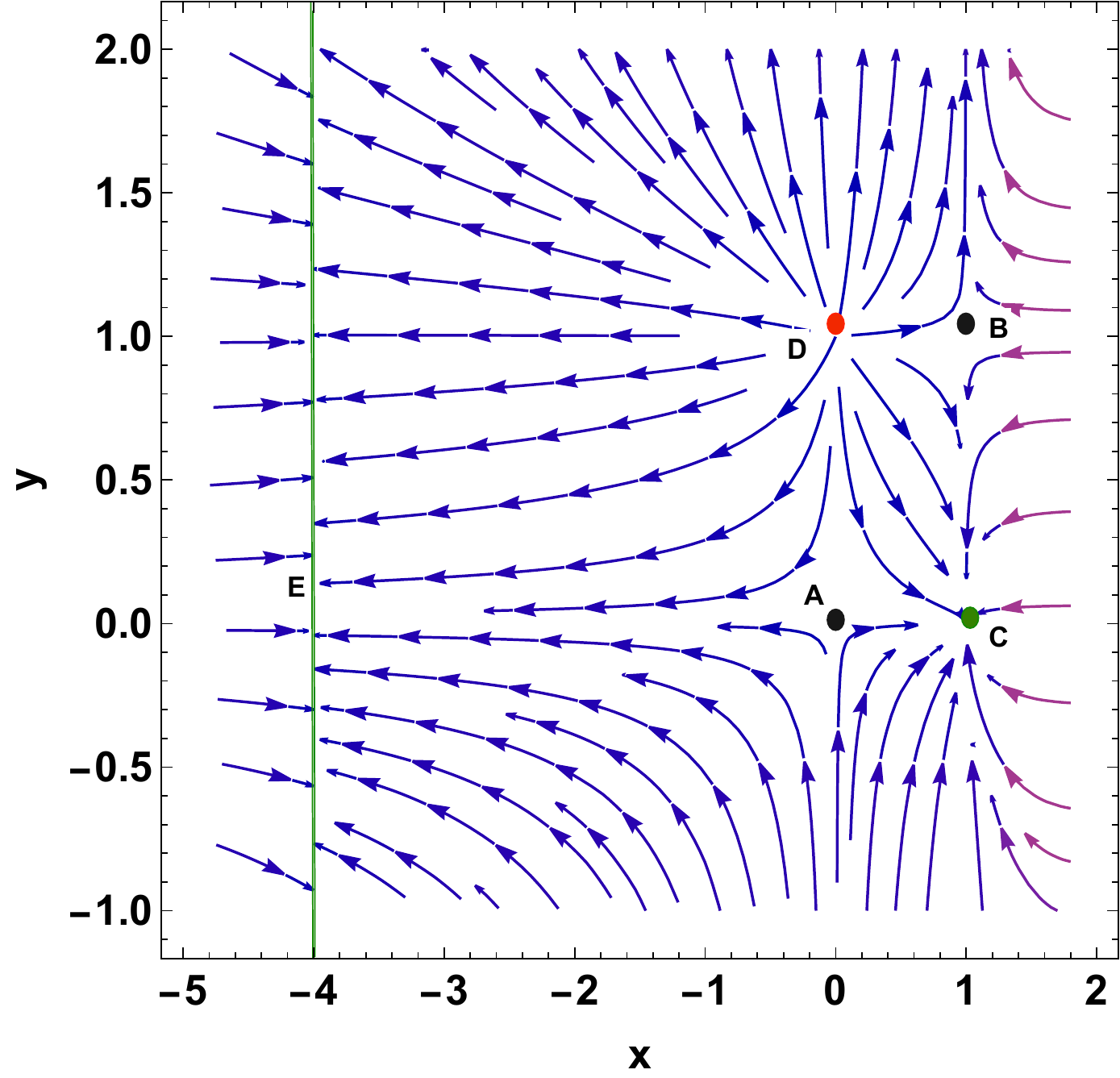}
\caption{Phase plot corresponding to the model $f(Q,T)=\alpha Q + \beta T^2 $ with $\kappa=1$. }\label{f2}
\end{center}
\end{figure}

\justify From Fig. \eqref{f2}, it is evident that the critical points A(0,0) and B(1,1) are saddle points, whereas the critical point C(1,0) indicates a stable matter dominated epoch. The critical point D(0,1) is a source while the critical point E(-4,y), for any y, is a sink representing a future attractor. The trajectories emerging from the D(0,1), indicating a decelerated epoch, are then converging to the point E(-4,y) representing an accelerated epoch. Thus we conclude that our non-linear f(Q,T) model efficiently describes the observed transition from decelerated phase to an accelerated phase of the universe. The considered model $f(Q,T)=\alpha Q+ \beta T^2$ with $\beta \neq 0$ behaves like standard $\Lambda$CDM model, and hence this model may represent an alternative to the $\Lambda$CDM.

 \subsubsection{Model-II}
We consider the following non-linear $f(Q,T)$ model
\begin{equation}\label{c0}
f(Q,T)= Q+ \alpha Q^2+ \beta T
\end{equation}
Here $\alpha \neq 0$ and $\beta$ are free model parameters. The astrophysical implications of the considered model have been investigated in \cite{Moreshwar}. We are going to examine the cosmological behavior of the model. 
The Friedmann equations \eqref{p} and \eqref{rho} corresponding to this non-linear $f(Q,T)$ model, for the dust case, becomes
\begin{equation}\label{c1}
    (2\kappa+3\beta ) \rho = 6 H^2 (1-18\alpha H^2)
\end{equation}
\begin{equation}\label{c2}
    2 \dot{H} (1-36\alpha H^2) + 3 H^2 (1-18\alpha H^2) = \frac{\beta}{2} \rho
\end{equation}
and the equation \eqref{dust} becomes
\begin{equation}\label{c3}
 \dot{\rho}  + \frac{6(\kappa+\beta)}{(2\kappa+3\beta)} H \rho = 0
\end{equation}
We define the following phase space variables 
\begin{equation}
    x=\frac{ (2\kappa+3\beta) \rho}{6H^2 (1-18\alpha H^2)}  \:\: \text{and} \:\: y= \frac{1}{\frac{H_0}{H}+1}
\end{equation}\label{c4}
Then we have constraints $x=1$ and $0 \leq y \leq 1 $ .
Now we define $N=dln(a)$, then corresponding to our non-linear $f(Q,T)$ model we obtained the following autonomous system by using \eqref{c1}-\eqref{c3},
\begin{equation}\label{c5}
    x'= \frac{dx}{dN} = \frac{3\beta}{(2\kappa+3\beta)} x (1-x)
\end{equation}
\begin{equation}\label{c6}
    y'= \frac{dy}{dN} = \frac{3(\kappa+\beta)}{(2\kappa+3\beta)}  \frac{[(1-y)^2-\bar{\alpha}y^2] }{[(1-y)^2-2 \bar{\alpha}y^2]} y (y-1)
\end{equation} 
Here $\bar{\alpha}=18\alpha H_0^2$.

Again, by the definition of deceleration and EoS parameter, we have 
\begin{equation}\label{c7}
    q= -1 + \frac{3 (\kappa+\beta ) [(1-y)^2-\bar{\alpha}  y^2]}{(2\kappa+3 \beta ) [(1-y)^2-2 \bar{\alpha}  y^2]}
\end{equation}
\begin{equation}\label{c8}
    \omega= -1 + \frac{2 (\kappa+\beta ) [(1-y)^2-\bar{\alpha}  y^2]}{(2\kappa+3 \beta ) [(1-y)^2-2 \bar{\alpha}  y^2]}
\end{equation}

 The nature of the critical points obtained by solving the above autonomous equations with $\kappa=1$, are presented below in the Table \eqref{Table-3}.

\begin{table}[H]
\begin{center}\caption{Table shows the critical points and their behavior corresponding to the model $f(Q,T)=Q+ \alpha Q^2+ \beta T$.}
\begin{tabular}{|c|c|c|c|c|c|}
\hline
 Critical Points $(x_c,y_c)$ & Eigenvalues $\lambda_1$ and $\lambda_2$ & Nature of critical point  & $q$ & $\omega$ \\ 
\hline 
 $A(1,0)$ & $-\frac{3(1+\beta)}{(2+3\beta)} \:\: \text{and} \:\: -\frac{3\beta}{(2+3\beta)}$ & Stable for $\beta < -1$ or $ \beta \geq 0$ & $\frac{1}{2+3\beta}$ & $-\frac{\beta}{2+3\beta}$ \\
  $B(1,1)$ & $-\frac{3\beta}{(2+3\beta)} \:\: \text{and} \:\: \frac{3(1+\beta)}{2(2+3\beta)}$ & Stable for $-1 < \beta < -\frac{2}{3}$  & $-\frac{1+3\beta}{2(2+3\beta)}$ & $-\frac{1+2\beta}{(2+3\beta)}$ \\
$C(1,\frac{1}{1+\sqrt{\bar{\alpha}}}), \bar{\alpha}>0$  & $-\frac{3\beta}{(2+3\beta)} \:\: \text{and} \:\: -\frac{6(1+\beta)}{(2+3\beta)}$ &  Stable for $\beta < -1$ or $ \beta \geq 0$ & $-1$ & $-1$ \\
\hline
\end{tabular}\label{Table-3}
\end{center}
\end{table}

The asymptotic behaviour of our non-linear $f(Q,T)$ model corresponding to the case $\beta=0$ with $\kappa=1$ and $\bar{\alpha}=1$, presented below in Fig. \eqref{f3}.

\begin{figure}[H]
\begin{center}
\includegraphics[scale=0.585]{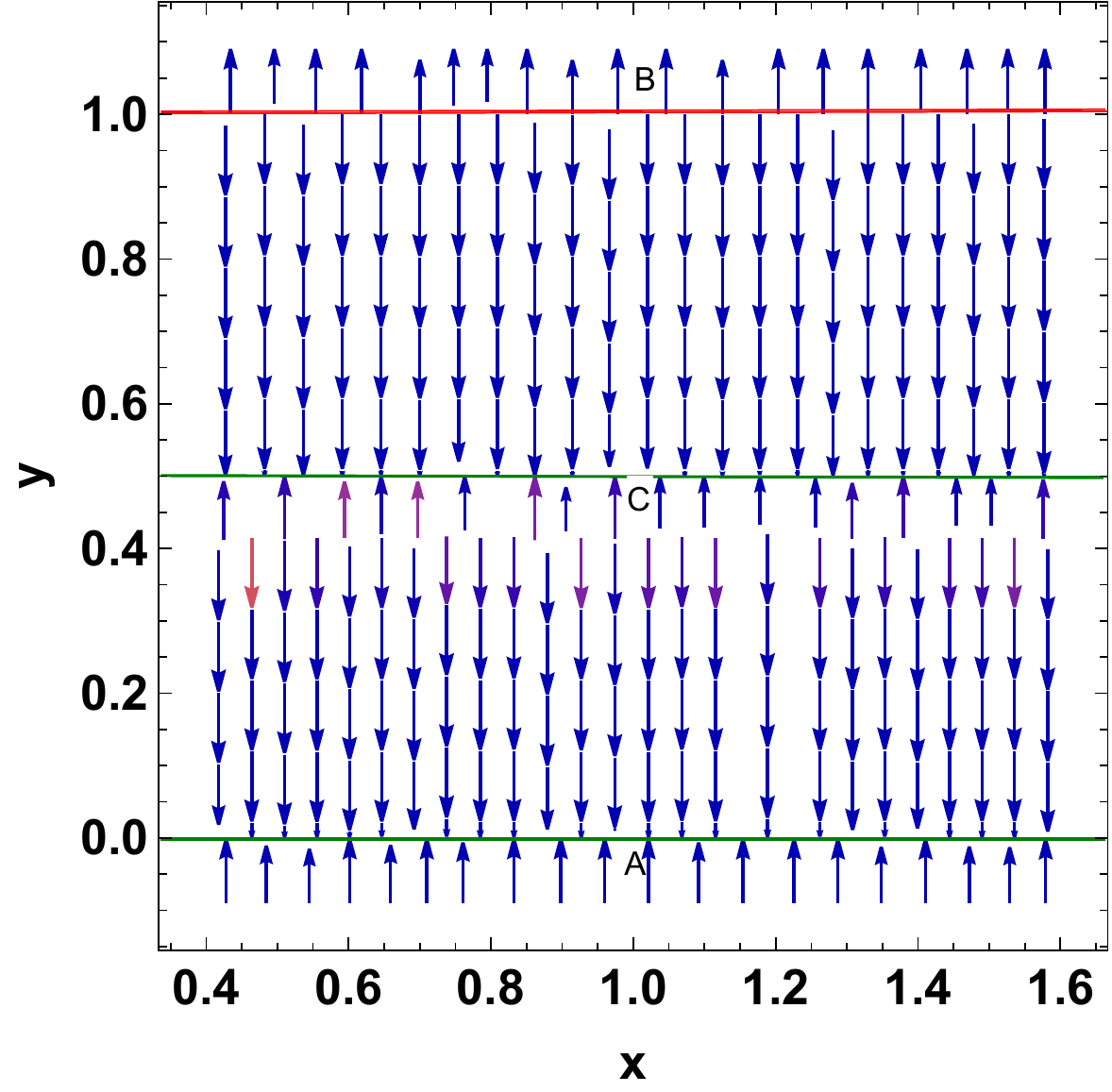}
\caption{Phase plots corresponding to the case $\beta=0$ with $\kappa=1$ and $\bar{\alpha}=1$. }\label{f3}
\end{center}
\end{figure}

For the case $\beta=0$ with $\kappa=1$ and $\bar{\alpha}=1$, the obtained critical points are A(1,0), B(1,1), and C(1,0.5). From Table \eqref{Table-3} and Fig. \eqref{f3}, it is evident that the critical points A(1,0) and C(1,0.5) with corresponding eigenvalues $-\frac{3}{2}, 0$ and $0,-3$ are stable and representing the matter dominated decelerated epoch and accelerated de-Sitter epoch respectively, whereas the point B(1,1) with eigenvalues $\frac{3}{4}, 0$ is unstable.   \\

%\textbf{Case II :}
%For the case $\beta=-1$ with $\kappa=1$ and $\bar{\alpha}=1$, we obtained non-isolated critical points R(1,y) with eigenvalues $0$ and $-3$. From the right panel of Fig. \eqref{f3}, it is evident that the trajectories are converging to the line of critical points R(1,y). Hence, R(1,y), for any y, is a sink representing the accelerated de-Sitter epoch.   \\

%%%%%%%%%%%%%%%%%%%%%%%%%%%%%%%%%%%%%%%%%%%%%%%%%%%%%%%%%%%%%%%%%
%%%%%%%%%%%%%%%%%%%%%%%%%%%%%%%%%%%%%%%%%%%%%%%%%%%%%%%%%%%%%%%%%

\section{Concluding remarks}\label{sec6}
In the present article we have examined the matter-geometry coupling in the form of the $f(Q,T)$ theory which shows promising development and rising research interest in the cosmological sector. We have derived an equivalent covariant formulation of the theory which yields an explicit comparison with the GR. The non-conservation of the energy-momentum tensor in this theory evokes a generally non-zero covariant vector which has been obtained from the divergence of this covariant formula of the field equation, and by using the Bianchi identity. In account of this result, we have computed the modified continuity relation, specific to this theory. The covariant formulation of the theory has also helped us to efficiently express the effective pressure and energy terms. Further, we have considered three different functional forms of the $f(Q,T)$ function, specifically, $f(Q,T)=\alpha Q+ \beta T$, $f(Q,T)=\alpha Q+ \beta T^2$, and $f(Q,T)=Q+ \alpha Q^2+ \beta T$ . These $f(Q,T)$ functions frequently appear in the literature. We have made an attempt to investigate the physical capabilities of the studied models to describe various cosmological epochs. We have incorporated the dynamical system technique to investigate the asymptotic behavior of the considered $f(Q,T)$ models. The obtained outcomes for the assumed linear and two non-linear $f(Q,T)$ models have been presented in Tables \eqref{Table-1}-\eqref{Table-3}, and the corresponding phase plots in Fig. \eqref{f1}-\eqref{f3}. Naturally, we have observed that the linear model $f(Q,T)=\alpha Q+ \beta T$ with $\beta=0$ is completely equivalent to the GR case without cosmological constant $\Lambda$. Further, we have found that the model $f(Q,T)=\alpha Q+ \beta T^2$ with $\beta \neq 0$ successfully describes the observed transition from decelerated phase to an accelerated phase of the universe and hence the considered non-linear model I behaves like standard $\Lambda$CDM model. Moreover, our obtained results agrees with the cosmological implications of the same model found in \cite{Yixin/2019}. Lastly, we found that the non-linear model II $f(Q,T)= Q+ \alpha Q^2+ \beta T$ with $\alpha \neq 0$ represents an accelerated de-Sitter epoch for the constraint $\beta < -1$ or $ \beta \geq 0$. Moreover, a stable matter dominated epoch with an accelerated de-Sitter type epoch can be obtained in the absence of energy momentum scalar term $T$ i.e., for the case $\beta=0$. From the basis of our findings we can conclude that both considered non-linear models can efficiently predicts the de-Sitter type expansion of the universe and may represent a viable geometric alternative to dark energy.\\
 
\textbf{Data availability:} There are no new data associated with this
article.\\

\textbf{Declaration of competing interest:} The authors declare that they
have no known competing financial interests or personal relationships that
could have appeared to influence the work reported in this paper.\\

%%%%%%%%%%%%%%%%%%%%%%%%%%%%%%%%%%%%%%%%%%%%%%%%%%%%%%%%%%%%%%%%%
%%%%%%%%%%%%%%%%%%%%%%%%%%%%%%%%%%%%%%%%%%%%%%%%%%%%%%%%%%%%%%%%%

\section*{Acknowledgments} 
The research was supported by the Ministry of Higher Education (MoHE), through the Fundamental Research Grant Scheme (FRGS/1/2021/STG06/UTAR/02/1). RS acknowledges UGC, New Delhi, India for providing Senior Research Fellowship with (UGC-Ref. No.: 191620096030). PKS  acknowledges the Science and Engineering Research Board, Department of Science and Technology, Government of India for financial support to carry out the Research project No.: CRG/2022/001847. 

%We are very much grateful to the honorable referee and to the editor for the
%illuminating suggestions that have significantly improved our work in terms of research quality, and presentation.

\end{document}